\documentclass[10pt,conference]{IEEEtran}
\pdfoutput=1
\usepackage{cite}
\usepackage{graphicx}
\usepackage{acronym}
\usepackage{algorithm}
\usepackage[noend]{algcompatible}
\usepackage[cmex10]{amsmath}
\usepackage[tight,footnotesize]{subfigure}
\usepackage{xspace}

\acrodef{QN}{Queueing Network}
\acrodef{QoS}{Quality of Service}
\acrodef{SLA}{Service Level Agreement}
\acrodef{BPEL}{Business Process Execution Language}
\acrodef{QN}{Queueing Network}
\acrodef{SAVER}{qoS-Aware workflows oVER the Cloud}
\acrodef{WS}{Web Service}
\acrodef{SaaS}{Software as a Service}
\acrodef{PaaS}{Platform as a Service}
\acrodef{IaaS}{Infrastructure as a Service}
\acrodef{VM}{Virtual Machine}

\newcommand{\curc}{\ensuremath{\mathbf{M}}} 
\newcommand{\curv}{\ensuremath{M}} 
\newcommand{\pippo}{\textsf{\ac{SAVER}}\xspace}

\newsavebox{\savefig}
\newenvironment{boxit}{\begin{lrbox}{\savefig}\begin{minipage}[b]{.9\columnwidth}}{\end{minipage}\end{lrbox}\fbox{\usebox{\savefig}}}

\newtheorem{lemma}{Lemma}

\begin{document}

\title{A Framework for QoS-aware Execution of Workflows over the Cloud}
\author{%
  \IEEEauthorblockN{Moreno Marzolla}
  \IEEEauthorblockA{Universit\`a di Bologna\\
    Dipartimento di Scienze dell'Informazione\\
    Mura A. Zamboni 7, I-40127 Bologna, Italy\\
    Email: marzolla@cs.unibo.it}
  \and
  \IEEEauthorblockN{Raffaela Mirandola}
  \IEEEauthorblockA{Politecnico di Milano\\
    Dipartimento di Elettronica e Informazione\\
    Piazza Leonardo da Vinci, I-20133 Milano, Italy\\
    Email: mirandola@elet.polimi.it}
}

\maketitle
\begin{abstract}
  The Cloud Computing paradigm is providing system architects with a
  new powerful tool for building scalable applications. Clouds allow
  allocation of resources on a "pay-as-you-go" model, so that
  additional resources can be requested during peak loads and released
  after that. However, this flexibility asks for appropriate dynamic
  reconfiguration strategies. In this paper we
  describe~\textsf{\acsu{SAVER}} (\acl{SAVER}), a QoS-aware algorithm
  for executing workflows involving Web Services hosted in a Cloud
  environment. \pippo allows execution of arbitrary workflows subject
  to response time constraints. \pippo uses a passive monitor to
  identify workload fluctuations based on the observed system response
  time. The information collected by the monitor is used by a planner
  component to identify the minimum number of instances of each Web
  Service which should be allocated in order to satisfy the response
  time constraint. \pippo uses a simple Queueing Network (QN) model to
  identify the optimal resource allocation. Specifically, the QN model
  is used to identify bottlenecks, and predict the system performance
  as Cloud resources are allocated or released. The parameters used to
  evaluate the model are those collected by the monitor, which means
  that \pippo does not require any particular knowledge of the Web
  Services and workflows being executed. Our approach has been
  validated through numerical simulations, whose results are reported
  in this paper.
\end{abstract}

\section{Introduction}\label{sec:intro}
The emerging Cloud computing paradigm is rapidly gaining consensus as
an alternative to traditional IT systems, as exemplified by the Amazon
EC2~\cite{ec2}, Xen~\cite{xen}, IBM Cloud~\cite{ibmcloud}, and
Microsoft Cloud~\cite{azure}. Informally, the Cloud computing paradigm
allows computing resources to be seen as a utility, available on
demand. The term ``resource'' may represent infrastructure, platforms,
software, services, or storage. In this vision, the Cloud provider is
responsible to make the resources available to the users as they
request it.

Cloud services can be grouped into three categories~\cite{cloudcomputing}:
\ac{IaaS}, providing low-level resources such as~\acp{VM}
(e.g., Amazon EC2~\cite{ec2});
\ac{PaaS}, providing software development frameworks
(e.g., Microsoft Azure~\cite{azure}); and~\ac{SaaS}, providing
applications (e.g., Salesforce.com~\cite{salesforce}). 

The Cloud provider has the responsibility to manage the resources it
provides (being them~\ac{VM} instances, programming frameworks or
applications) so that the user requirements and the desired~\ac{QoS}
are satisfied. Cloud users are usually charged according to the amount
of resources they consume (e.g., some amount of money per hour of CPU
usage). In this way, customers can avoid capital expenditures by using
Cloud resources on a ``pay-as-you-go'' model.

Users~\ac{QoS} requirements (e.g., timeliness, availability, security)
are usually the result of a negotiation process engaged between the
resource provider and the user, which culminates in the definition of
a~\ac{SLA} concerning their respective obligations and expectations.
Guaranteeing~\acp{SLA} under variable workloads for different
application and service models is extremely challenging: Clouds are
characterized by high load variance, and users have heterogeneous and
competing~\ac{QoS} requirements.

In this paper we present~\textsf{\acs{SAVER}} (\acl{SAVER}), a
workflow engine provided as a~\ac{SaaS}.  The engine allows different
types of workflows to be executed over a set of~\acp{WS}. Workflows
are described using some appropriate notations (e.g., using the
WS-BPEL~\cite{bpel} workflow description language). The workflow
engine takes care of interacting with the appropriate~\acp{WS} as
described in the workflow.

In our scenario, users can negotiate~\ac{QoS} requirements with the
service provider; specifically, for each type $c$ of workflow, the
user may request that the average execution time of the whole workflow
should not exceed a threshold $R_c^+$. Once the~\ac{QoS} requirements
have been negotiated, the user can submit any number of workflows of
the different types. Both the submission rate and the time spent by
the workflows on each~\ac{WS} can fluctuate over time.

Traditionally, when deciding the amount of resources to be dedicated
to applications, service providers considered worst-case scenarios,
resulting in resource over-provisioning. Since the worst-case scenario
rarely happens, a static system deployment results in a processing
infrastructure which is largely under-utilized.

To increase the utilization of resources while meeting the
requested~\ac{SLA}, \pippo uses an underlying~\ac{IaaS} Cloud to
provide computational power on demand. The Cloud hosts multiple
instances of each~\ac{WS}, so that the workload can be balanced across
the instances. If a~\ac{WS} is heavily used, \pippo will increase the
number of instances by requesting new resources from the Cloud. In
this way, the response time of that~\ac{WS} can be reduced, reducing
the total execution time of workflows as well. \pippo monitors the
workflow engine and detects when some constraints are being
violated. System reconfigurations are triggered periodically, when
instances are added or removed where necessary.

\begin{figure}[t]
\centering\includegraphics[width=\columnwidth]{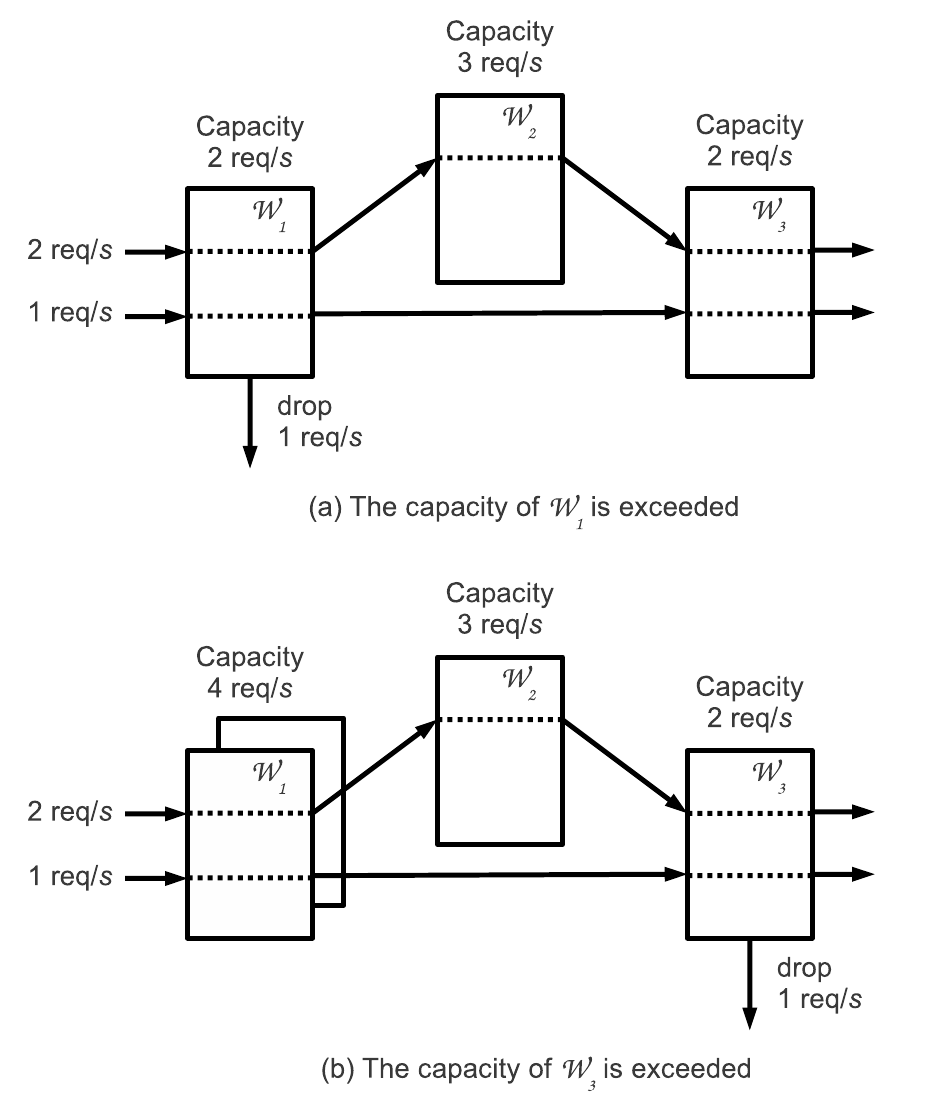}
\caption{Illustration of the bottleneck shift issue}\label{fig:example}
\end{figure}

Despite its conceptual simplicity, the idea above is quite challenging
to implement in practice. To better illustrate the problem, let us
consider the situation shown in Fig,~\ref{fig:example}, which is
modeled upon a similar example from~\cite{urgaonkar}. We have three
Web Services $\mathcal{W}_1, \mathcal{W}_2, \mathcal{W}_3$ which are
used by two types of workflows. Instances of the first type arrive at
a rate of 2 $\mathit{req}/s$, and execute operations on
$\mathcal{W}_1$, $\mathcal{W}_2$ and $\mathcal{W}_3$.  Instances of
the second workflow type arrive at a rate of 1 $\mathit{req}/s$ and
only use $\mathcal{W}_1$ and $\mathcal{W}_3$. Each~\ac{WS} has a
maximum capacity, which corresponds to the maximum request rate it can
handle. Web Services 1 and 3 have a maximum capacity of 2
$\mathit{req}/s$, while~\ac{WS} 2 has a capacity of 3
$\mathit{req}/s$.

In Fig.~\ref{fig:example}(a) the capacity of $\mathcal{W}_1$ is
exceeded, because the aggregate arrival rate (3 $\mathit{req}/s$) is
greater than its processing capacity. Thus, a queue of unprocessed
invocations of $\mathcal{W}_1$ builds up, until requests start to
timeout and are dropped at a rate of 1 $\mathit{req}/s$. To eliminate
the bottleneck, a possible solution is to create multiple instances of
the bottleneck~\ac{WS} on different servers, and balance the load
across all instances. If we apply this strategy and create two
instances of $\mathcal{W}_1$, we get the situation shown in
Fig.~\ref{fig:example}(b): the aggregate processing capacity of
$\mathcal{W}_1$ is now 4 $\mathit{req}/s$, and thus Web Service 1 is
no longer the bottleneck. However, the bottleneck shifts to
$\mathcal{W}_3$, which now sees an aggregate arrival rate of 3
$\mathit{req}/s$ and has a capacity of 2 $\mathit{req}/s$.

The situation above demonstrates the
\emph{bottleneck shift} phenomenon: fixing a bottleneck may create
another bottleneck at a different place. Thus, satisfying QoS
constraints on systems subject to variable workloads is challenging,
because identifying the system configuration which satisfies all
constraints might involve multiple reconfigurations of individual
components (in our scenario, adding~\ac{WS} instances). If the
reconfiguration is implemented in a purely reactive manner, each step
must be applied sequentially in order to monitor its impact and plan
for the next step. This is clearly inefficient because adaptation
would be exceedingly slow.

In general, the response time at a specific~\ac{WS} depends both on
the number of instances of that Web Service, and also on the intensity
of other workload classes (workflow types). Thus, a suitable system
performance model must be used in order to predict the response time
of a given configuration. The performance model can be used to drive
the reconfiguration process proactively: different system
configurations can be evaluated quickly, and multiple reconfiguration
steps can be planned in advance. \pippo uses a open,
multiclass~\ac{QN} model to represent resource contention by multiple
independent request flows, which is crucial in our scenario.  The
parameters which are needed to evaluate the~\ac{QN} model can be
easily obtained by passively monitoring the running system. The
performance model is used within a greedy strategy which identifies an
approximate solution to the optimization problem minimizing the number
of~\ac{WS} instances while respecting the~\ac{SLA}.

\paragraph*{Structure of this paper}
The remainder of this paper is organized as follows. In
Section~\ref{sec:related_works} we review the scientific literature
and compare \pippo with related works. In
Section~\ref{sec:problem_formulation} we give a precise formulation of
the problem we are addressing. In Section~\ref{sec:perfmodel} we
describe the Queueing Network performance model of the Cloud-based
workflow engine. \pippo will be fully described in
Section~\ref{sec:architecture}, including the high-level architecture
and the details of the reconfiguration algorithms. The effectiveness
of \pippo have been evaluated by means of simulation experiments,
whose results will be discussed in Section~\ref{sec:results}. Finally,
conclusions and future works are presented in
Section~\ref{sec:conclusions}. In order to make this paper
self-contained without sacrificing clarify, we relegated the
mathematical details of the analysis of the performance model in a
separate Appendix.

\section{Related works}\label{sec:related_works}

Several research contributions have previously addressed the issue of
optimizing the resource allocation in cluster-based service
centers. Recently, with the emerging of virtualization approaches and
Cloud computing, additional research on automatic resource management
has been conducted. In this section we briefly review some recent
results; some of them take advantage of control theory-based feedback
loops~\cite{Litoiu2010,Kalyvianaki09}, machine learning
techniques~\cite{Kephart07,Calinescu09}, or utility-based optimization
techniques~\cite{Urgaonkar07,Zhu09}.
 
When moving to virtualized environments the resource allocation
problem becomes even more complex because of the introduction of
virtual resources~\cite{Zhu09}. Several approaches have been proposed
for~\ac{QoS} and resource management at
run-time~\cite{Litoiu2009,Litoiu2010,mistral2010,Ferretti2010,kounev2011,Yazir2010}.

The approach presented in~\cite{Litoiu2009} describes a method for
achieving optimization in Clouds by using performance models all along
the development and operation of the applications running in the
Cloud. The proposed optimization aims at maximizing profits in the
Cloud by guaranteeing the QoS agreed in the~\acp{SLA} taking into
account a large variety of workloads.  A layered Cloud architecture
taking into account different stakeholders is presented
in~\cite{Litoiu2010}. The architecture supports self-management based
on adaptive feedback control loops, present at each layer, and on a
coordination activity between the different loops.
Mistral~\cite{mistral2010} is a resource managing framework with a
multi-level resource allocation algorithm considering reallocation
actions based mainly on adding, removing and/or migrating virtual
machines, and shutdown or restart of hosts. This approach is based on
the usage of Layered Queuing Network (LQN) performance model. It tries
to maximize the overall utility taking into account several aspects
like power consumption, performance and transient costs in its
reconfiguration process.  In~\cite{kounev2011} the authors present an
approach to self-adaptive resource allocation in virtualized
environments based on online architecture-level performance models.
The online performance prediction allow estimation of the effects
of changes in user workloads and of possible reconfiguration actions.
Yazir {\em et al.}~\cite{Yazir2010} introduces a distributed approach
for dynamic autonomous resource management in computing Clouds,
performing resource configuration using through Multiple Criteria
Decision Analysis.

With respect to these works, \pippo lies in the same research line
fostering the usage of models at runtime to drive the QoS-based system
adaptation. \pippo uses an efficient modeling and analysis technique
that can then be used at runtime without undermining the system
behavior and its overall performance.

Ferretti {\em et al.} propose in~\cite{Ferretti2010} a middleware
architecture enabling a SLA-driven dynamic configuration, management
and optimization of Cloud resources and services. The approach makes
use of a load balancer that distributes the workload among the
available resources. When the perceived QoS deviates from the SLA, the
platform is dynamically reconfigured by acquiring new resources from
the Cloud. On the other hand, if resources under-utilization is
detected, the system triggers a reconfiguration to release those
unused resources. This approach is purely reactive and considers a
single-tier application, while \pippo works for an arbitrary number
of~\acp{WS} and uses a performance model to plan complex
reconfigurations in a single step.

Canfora {\em et al.}~\cite{Canfora2005} describe a~\ac{QoS}-aware
service discovery and late-binding mechanism which is able to
automatically adapt to changes of~\ac{QoS} attributes in order to meet
the~\ac{SLA}. The authors consider the execution of workflows over a
set of~\acp{WS}, such that each~\ac{WS} has multiple functionally
equivalent implementations. Genetic Algorithms are use to bind
each~\ac{WS} to one of the available implementations, so that a
fitness function is maximized. The binding is done at run-time, and
depends on the values of~\ac{QoS} attributes which are monitored by
the system. It should be observed that in \pippo we consider a
different scenario, in which each~\ac{WS} has just one implementation
which however can be instantiated multiple times. The goal of \pippo
is to satisfy a specific~\ac{QoS} requirement (mean execution time of
workflows below a given threshold) with the minimum number of
instances.

\section{Problem Formulation}\label{sec:problem_formulation}

\begin{figure}
  \centering\includegraphics[width=\columnwidth]{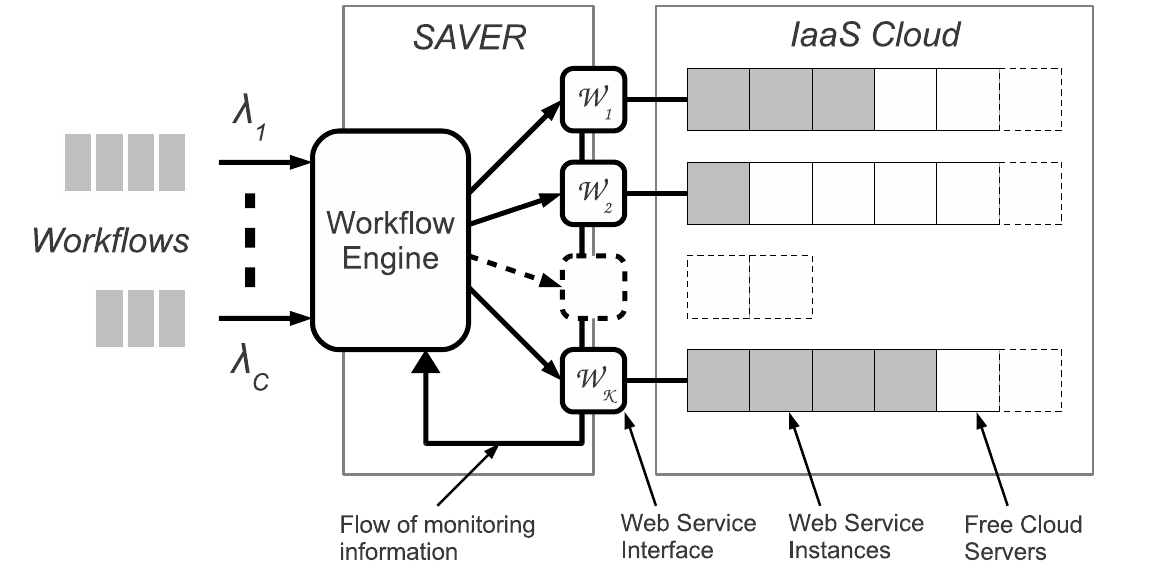}
  \caption{System model}\label{fig:system}
\end{figure}

\pippo is a workflow engine whose general structure is depicted in
Fig.~\ref{fig:system}: it receives workflows from external clients,
and executes them over a set of $K$~\ac{WS} $\mathcal{W}_1, \ldots,
\mathcal{W}_K$.  Workflows can be of $C$ different types (or classes);
for each class $c=1, \ldots, C$, clients define a maximum allowed
completion time $R_c^+$. This means that an instance of class $c$
workflow must be completed, on average, in time less than
$R_c^+$. New workflow classes can be created at any time; when a new
class is created, its maximum response time is negotiated with the
workflow service provider.

We denote with $\lambda_c$ the average arrival rate of class $c$
workflows. Arrival rates can change over time\footnote{In order to
  simplify the notation, we write $\lambda_c$ instead of
  $\lambda_c(t)$. In general, we will omit explicit reference to $t$
  for all time-dependent parameters.}.  Since all~\acp{WS} are shared
between the workflows, the completion time of a workflow depends both
on arrival rates $\boldsymbol{\lambda} = (\lambda_1, \ldots,
\lambda_C)$, and on the utilization of each~\ac{WS}.

In order to satisfy the response time constraints, the system must
adapt to cope with fluctuations of the workload. To do so, \pippo
relies on a~\ac{IaaS} Cloud which maintains multiple instances of
each~\ac{WS}. Run-time monitoring information is sent by all~\acp{WS}
back to the workflow engine to drive the adaptation process.  We
denote with $N_k$ the number of instances of~\ac{WS} $\mathcal{W}_k$;
a system configuration $\mathbf{N} = (N_1, \ldots, N_K)$ is an integer
vector representing the number of allocated instances of each~\ac{WS}.

When a workflow interacts with $\mathcal{W}_k$, it is bound to one of
the $N_k$ instances so that the requests are evenly distributed. When
the workload intensity increases, additional instances are created to
eliminate the bottlenecks; when the workload decreases, surplus
instances are shut down and released.

The goal of~\pippo is to minimize the total number of~\ac{WS}
instances while maintaining the mean execution time of type $c$
workflows below the threshold $R_c^+$, $c=1, \ldots, C$. Formally, we
want to solve the following optimization problem:

\begin{eqnarray}
\textrm{minimize} && f(\mathbf{N}) = \sum_{k=1}^K N_k\label{eq:problem}\\
\textrm{subject to} && R_c(\mathbf{N}) \leq R_c^+ \qquad \mbox{for all}\ c=1, 2, \ldots, C\nonumber\\
&& N_i \in \{1, 2, 3, \ldots\}\nonumber
\end{eqnarray}

\noindent where $R_c(\mathbf{N})$ is the mean execution time of type
$c$ workflows when the system configuration is $\mathbf{N} = (N_1,
\ldots, N_K)$.

If the~\ac{IaaS} Cloud which hosts~\ac{WS} instances is managed by
some third-party organization, then reducing the number of active
instances reduces the cost of the workflow engine.

\section{System Performance Model}\label{sec:perfmodel}

\begin{figure}[t]
\centering\includegraphics[width=\columnwidth]{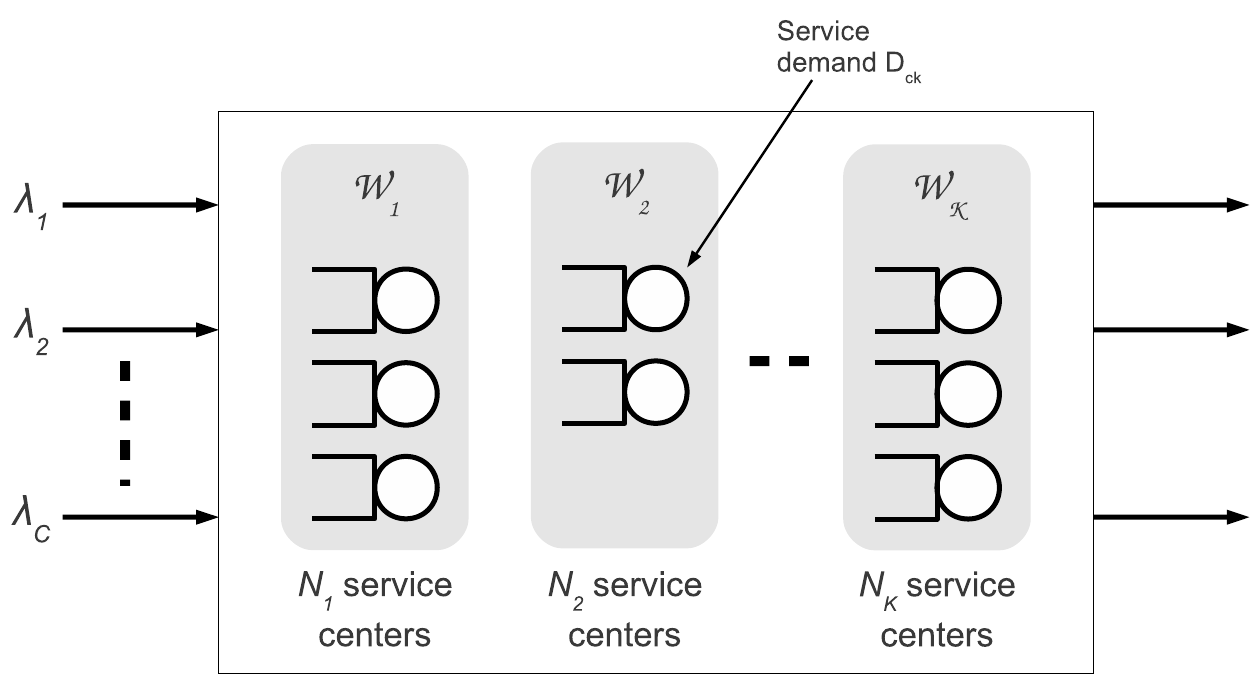}
\caption{Performance model based on an open, multiclass Queueing Network}\label{fig:model}
\end{figure}

Before illustrating the details of \pippo, it is important to describe
the~\ac{QN} performance model which is used to plan a system
reconfiguration. We model the system of Fig.~\ref{fig:system} using
the open, multiclass~\ac{QN} model~\cite{lazowska} shown in
Fig.~\ref{fig:model}. A~\ac{QN} model is a set of queueing centers,
which in our case are FIFO queues attached to a single server. Each
server represents a single~\ac{WS} instance; thus, $\mathcal{W}_k$ is
represented by $N_k$ queueing centers, for each $k=1, \ldots, K$.
$N_k$ can change over time, as resources are added or removed from the
system.

In our~\ac{QN} model there are $C$ different classes of requests,
which are generated outside the system. Each request represents a
workflow, thus workflow types are directly mapped to~\ac{QN} request
classes. In order to simplify the analysis of the model, we make the
simplifying assumption that the inter-arrival time of class $c$
requests is exponentially distributed with arrival rate
$\lambda_c$. This means that a new workflow of type $c$ is submitted,
on average, every $1 / \lambda_c$ time units.

The interaction of a type $c$ workflow with~\ac{WS} $\mathcal{W}_k$ is
modeled as a visit of a class $c$ request to one of the $N_k$ queueing
centers representing $\mathcal{W}_k$. We denote with $R_{c
  k}(\mathbf{N})$ the total time (\emph{residence time}) spent by type
$c$ workflows on one of the $N_k$ instances of $\mathcal{W}_k$ for a
given configuration $\mathbf{N}$.  The residence time is the sum of
two terms: the \emph{service demand} $D_{c k}(\mathbf{N})$ (average
time spent by a~\ac{WS} instance executing the request) and queueing
delay (time spent by a request in the waiting queue). The~\ac{QN}
model allows multiple visits to the same queueing center, because the
same~\ac{WS} can be executed multiple times by the same workflow. The
residence time and service demands are the sum of residence and
service time of \emph{all} invocations of the same~\ac{WS} instance.

The \emph{utilization} $U_k(\mathbf{N})$ of an instance of
$\mathcal{W}_k$ is the fraction of time the instance is busy
processing requests. If the workload is evenly balanced, then both the
residence time $R_{c k}(\mathbf{N})$ and the utilization
$U_k(\mathbf{N})$ are almost the same for all $N_k$ instances of
$\mathcal{W}_k$.

Table~\ref{tab:symbols} summarizes the symbols used in this paper.

\begin{table}[t]
\caption{Symbols used in this paper}\label{tab:symbols}%
\centering%
\begin{tabular}{ll}
\hline
$C$ & Number of workflow types\\
$K$ & Number of Web Services\\
$\boldsymbol{\lambda}$ & Vector of per-class Arrival rates\\
$\curc$ & Current system configuration\\
$\mathbf{N}, \mathbf{N}'$ & Arbitrary system configurations\\
$R_{c k}(\mathbf{N})$ & Residence time of type $c$ workflows on an instance of $\mathcal{W}_k$\\
$D_{c k}(\mathbf{N})$ & Service demand of type $c$ workflows on an instance of $\mathcal{W}_k$\\
$R_c(\mathbf{N})$ & Response time of type $c$ workflows\\
$U_k(\mathbf{N})$ & Utilization of an instance of $\mathcal{W}_k$\\
$R_c^+$ & Maximum allowed response time for type $c$ workflows\\
\hline
\end{tabular}
\end{table}

\section{Architectural Overview of \pippo}\label{sec:architecture}

\begin{figure}[t]
\includegraphics[width=\columnwidth]{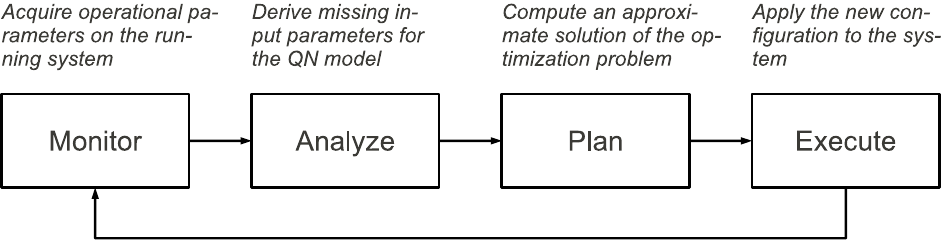}
\caption{\pippo Control Loop}\label{fig:control}
\end{figure}

\pippo is a reactive system based on the Monitor-Analyze-Plan-Execute
(MAPE) control loop shown in Fig.~\ref{fig:control}. During the
\emph{Monitor} step, \pippo collects operational parameters by
observing the running system. The parameters are evaluate during the
\emph{Analyze} step; if the system needs to be reconfigured (e.g.,
because the observed response time of class $c$ workflows exceeds the
threshold $R_c^+$, for some $c$), a new configuration is identified in
the \emph{Plan} step. We use the~\ac{QN} model described in
Section~\ref{sec:perfmodel} to evaluate different configurations and
identify an optimal server allocation such that all~\ac{QoS}
constraints are satisfied. Finally, during the \emph{Execute} step,
the new configuration is applied to the system: \ac{WS} instances are
created or destroyed as needed by leveraging the~\ac{IaaS} Cloud.
Unlike other reactive systems, \pippo can plan complex
reconfigurations, involving multiple additions/removals of resources,
in a single step.

\subsection{Monitoring System Parameters}\label{sec:monitoring}

The~\ac{QN} model is used to estimate the execution time of workflow
types for different system configurations. To analyze the~\ac{QN} it
is necessary to know two parameters: (\emph{i}) the arrival rate of
type $c$ workflows, $\lambda_c$, and (\emph{ii}) the service demand
$D_{c k}(\curc)$ of type $c$ workflows on an instance of~\ac{WS}
$\mathcal{W}_k$, for the current configuration $\curc$.

The parameters above can be computed by monitoring the system over a
suitable period of time. The arrival rates $\lambda_c$ can be
estimated by counting the number $A_c$ or arrivals of type $c$
workflows which are submitted over the observation period of length
$T$. Then $\lambda_c$ can be defined as $\lambda_c = A_c / T$.

\begin{table}[t]
\caption{Equations for the~\ac{QN} model of Fig.~\ref{fig:model}}\label{tab:eq}
\centering%
\begin{boxit}%
\begin{eqnarray}%
U_k(\mathbf{N}) &=& \sum_{c=1}^C \lambda_c D_{c k}(\mathbf{N}) \label{eq:utilization} \\
R_{c k}(\mathbf{N}) &=& \frac{D_{c k}(\mathbf{N})}{1 - U_k(\mathbf{N})}\label{eq:resptimeck}\\
R_c(\mathbf{N}) &=& \sum_{k=1}^K N_k R_{c k}(\mathbf{N})\label{eq:resptime}
\end{eqnarray}
\end{boxit}
\end{table}

Measuring the service demands $D_{c k}(\curc)$ is a bit more difficult
because they must not include the time spent by a request waiting to
start service.  If the~\acp{WS} do not provide detailed timing
information (e.g., via their execution logs), it is possible to
estimate $D_{c k}(\curc)$ from parameters which can be easily observed
by the workflow engine, that are the measured residence time $R_{c
  k}(\curc)$ and utilization $U_k(\curc)$. We use the equations shown
in Table~\ref{tab:eq}, which hold for the open multiclass~\ac{QN}
model in Fig.~\ref{fig:model}. These equations describe well known
properties of open~\ac{QN} models, so they are given here without any
proof. The interested reader is referred to~\cite{lazowska} for
details.

The residence time is the total time spent by a type $c$ workflow with
one instance of~\ac{WS} $\mathcal{W}_k$, including waiting time and
service time. The workflow engine can measure $R_{c k}(\curc)$ as the
time elapsed from the instant a type $c$ workflow sends a request to
one of the $N_k$ instances of $\mathcal{W}_k$, to the time the request
is completed. The utilization $U_k(\curc)$ of an instance of
$\mathcal{W}_k$ can be obtained by the Cloud service dashboard (or
measured on the computing nodes
themselves). Using~\eqref{eq:resptimeck} the service demands can be
expressed as
\begin{equation}
D_{c k}(\curc) = R_{c k}(\curc) \left(1 - U_k(\curc) \right)\label{eq:estdemand}
\end{equation}

\subsection{Finding a new configuration}

\begin{algorithm}[t]
\caption{The \pippo Algorithm}\label{alg:pippo}
\begin{algorithmic}[1]
\REQUIRE{$R_c^+$: Maximum response time of type $c$ workflows}
\STATE{Let $\mathbf{M}$ be the initial configuration}\label{line:initconf}
\LOOP
\STATE{Monitor $R_{c k}(\curc)$, $U_k(\curc)$, $\lambda_c$}
\FORALL{$c:=1, \ldots, C;\ k:=1, \ldots, K$}
\STATE{Compute $D_{c k}(\curc)$ using Eq.~\eqref{eq:estdemand}}
\ENDFOR
\STATE{$\mathbf{N} := \textsf{Acquire}(\curc, \boldsymbol{\lambda}, \mathbf{D}(\curc), \mathbf{U}(\curc) )$}
\FORALL{$c:=1, \ldots, C;\ k:=1, \ldots, K$}
\STATE{Compute $D_{c k}(\mathbf{N})$ and $U_k(\mathbf{N})$ using Eq.~\eqref{eq:demand1} and~\eqref{eq:utilization1}}
\ENDFOR
\STATE{$\mathbf{N}' := \textsf{Release}(\mathbf{N}, \boldsymbol{\lambda}, \mathbf{D}(\mathbf{N}), \mathbf{U}(\mathbf{N}) )$}
\STATE{Apply the new configuration $\mathbf{N}'$ to the system}
\STATE{$\curc := \mathbf{N}'$}\hfill\COMMENT{Set $\mathbf{N}'$ as the current configuration \curc}
\ENDLOOP
\end{algorithmic}
\end{algorithm}

In order to find an approximate solution to the optimization
problem~\eqref{eq:problem}, \pippo starts from the current
configuration $\curc$, which may violate some response time
constraints, and executes Algorithm~\ref{alg:pippo}. After collecting
device utilizations, response times and arrival rates, \pippo
estimates the service demands $D_{c k}$ using
Eq.~\eqref{eq:estdemand}.

Then, \pippo identifies a new configuration $\mathbf{N} \geq
\curc$\footnote{$\mathbf{N} \geq \mathbf{M}$ iff $N_k \geq M_k$ for
  all $k=1, \ldots, K$} by calling the function
\textsc{Acquire}(). The new configuration $\mathbf{N}$ is computed by
greedily adding new instances to bottleneck~\acp{WS}. The~\ac{QN}
model is used to estimate response times as instances are added: no
actual resources are instantiated from the Cloud service at this time.

The configuration $\mathbf{N}$ returned by the function
\textsc{Acquire}() does not violate any constraint, but might contain
too many~\ac{WS} instances. Thus, \pippo invokes the function
\textsc{Release}() which computes another configuration $\mathbf{N}'
\leq \mathbf{N}$ by removing redundant instances, ensuring that no
constraint is violated. To call procedure \textsc{Release}() we need
to estimate the service demands $D_{ck}(\mathbf{N})$ and utilizations
$U_k(\mathbf{N})$ with configuration $\mathbf{N}$. These can be easily
computed from the measured values for the current configuration
$\curc$.

After both steps above, $\mathbf{N}'$ becomes the new current
configuration: \ac{WS} instances are created or terminated where
necessary by acquiring or releasing hosts from the Cloud
infrastructure.

Let us illustrate the functions \textsc{Acquire}() and
\textsc{Release}() in detail.

\paragraph{Adding instances} 
Function \textsc{Acquire}() is described by
Algorithm~\ref{alg:acquire}. Given the system parameters and
configuration $\mathbf{N}$, which might violate some or all response
time constraints, the function returns a new configuration
$\mathbf{N}'$ which is estimated not to violate any constraint. At
each iteration, we identify the class $b$ whose workflows have the
maximum relative violation of the response time limit
(line~\ref{line:estb}); response times are estimated using
Eq.~\eqref{eq:respestimate} in the Appendix. Then, we identify
the~\ac{WS} $\mathcal{W}_j$ such that adding one more instance to it
produces the maximum reduction in the class $b$ response time
(line~\ref{line:estj}). The configuration $\mathbf{N}$ is then updated
by adding one instance to $\mathcal{W}_j$ (line~\ref{line:incrN}); the
updated configuration is $\mathbf{N} +
\mathbf{1}_j$\footnote{$\mathbf{1}_j$ is a vector with $K$ elements,
  whose $j$-th element is one and all others are set to zero}. The
loop terminates when no workload type is estimated to violate its
response time constraint.

\begin{algorithm}[t]
\caption{\textsf{Acquire}$(\mathbf{N}, \boldsymbol{\lambda}, \mathbf{D}(\mathbf{N}), \mathbf{U}(\mathbf{N})) \rightarrow \mathbf{N}'$}\label{alg:acquire}
\begin{small}
\begin{algorithmic}[1]
\REQUIRE{$\mathbf{N}$ System configuration}
\REQUIRE{$\boldsymbol{\lambda}$ Current arrival rates of workflows}
\REQUIRE{$\mathbf{D}(\mathbf{N})$ Service demands at configuration $\mathbf{N}$}
\REQUIRE{$\mathbf{U}(\mathbf{N})$ Utilizations at configuration $\mathbf{N}$}
\ENSURE{$\mathbf{N}$ New system configuration}
\WHILE{$\left( R_c(\mathbf{N}) > R_c^+\ \mbox{for any}\ c\right)$}\smallskip
\STATE{$\displaystyle b := \arg\max_c \left\{\left. \frac{R_c(\mathbf{N}) - R_c^+}{R_c^+}\ \right|\ c=1, \ldots, C \right\}$}\label{line:estb}\smallskip
\STATE{$\displaystyle j := \arg\max_k \left\{ \left. R_b(\mathbf{N}) - R_b(\mathbf{N}+\mathbf{1}_k)\ \right|\ k=1, \ldots, K \right\}$}\label{line:estj}\smallskip
\STATE{$\mathbf{N} := \mathbf{N} + \mathbf{1}_j$}\label{line:incrN}
\ENDWHILE
\STATE{\textbf{Return} $\mathbf{N}$}
\end{algorithmic}
\end{small}
\end{algorithm}

Termination of Algorithm~\ref{alg:acquire} is guaranteed by the fact
that function $R_c(\mathbf{N})$ is monotonically decreasing
(Lemma~\ref{lemma:monotonically} in the Appendix). Thus,
$R_c(\mathbf{N}+\mathbf{1}_j) < R_c(\mathbf{N})$ for all $c$.

\paragraph{Removing instances}
The function \textsc{Release}(), described by
Algorithm~\ref{alg:release}, is used to deallocate (release)~\ac{WS}
instances from an initial configuration $\mathbf{N}$ which does not
violate any response time constraint. The function implements a greedy
strategy, in which a~\ac{WS} $\mathcal{W}_j$ is selected at each step,
and its number of instances is reduced by one. Reducing the number of
instances $N_j$ of $\mathcal{W}_j$ is not possible if, either
(\emph{i}) the reduction would violate some constraint, or (\emph{ii})
the reduction would cause the utilization of some~\ac{WS} instances to
become greater than one (see Eq.~\eqref{eq:capconstraint} in the
Appendix).

We start by defining the set $S$ containing the index of~\acp{WS}
whose number of instances can be reduced without exceeding the
processing capacity (line~\ref{line:defS}). Then, we identify the
workflow class $d$ with the maximum (relative) response time
(line~\ref{line:defd}). Finally, we identify the value $j \in S$ such
that removing one instance of $\mathbf{W}_j$ produces the minimum
increase in the response time of class $d$ workflows
(line~\ref{line:defj}). The rationale is the following. Type $d$
workflows are the most likely to be affected by the removal of
one~\ac{WS} instance, because their relative response time (before the
removal) is the highest among all workflow types. Once the
``critical'' class $d$ has been identified, we try to remove an
instance from the~\ac{WS} $j$ which causes the smallest increase of
class $d$ response time. Since response time increments are additive
(see Appendix), if the removal of an instance of $\mathcal{W}_j$
violates some constraints, no further attempt should be done to
consider $\mathcal{W}_j$, and we remove $j$ from the candidate set
$S$.

From the discussion above, we observe that function \textsc{Release}()
computes a \emph{Pareto-optimal} solution $\mathbf{N}$. This means
that there exists no solution $\mathbf{N}' \leq \mathbf{N}$ such that
$R_c(\mathbf{N}') \leq R_c^+$.

\begin{algorithm}[t]
\caption{\textsf{Release}$(\mathbf{N}, \boldsymbol{\lambda}, \mathbf{D}(\mathbf{N}), \mathbf{U}(\mathbf{N})) \rightarrow \mathbf{N}'$}\label{alg:release}
\begin{small}
\begin{algorithmic}[1]
\REQUIRE{$\mathbf{N}$ System configuration}
\REQUIRE{$\boldsymbol{\lambda}$ Current arrival rates of workflows}
\REQUIRE{$\mathbf{D}(\mathbf{N})$ Service demands at configuration $\mathbf{N}$}
\REQUIRE{$\mathbf{U}(\mathbf{N})$ Utilizations at configuration $\mathbf{N}$}
\ENSURE{$\mathbf{N}'$ New system configuration}

\FORALL{$k := 1, \ldots, K$}
\STATE{$\mathit{Nmin}_k := N_k \sum_{c=1}^C \lambda_c D_{c k}(\mathbf{N})$}
\ENDFOR
\STATE{$S := \left\{k\ |\ N_k > \mathit{Nmin}_k\right\}$}\label{line:defS}
\WHILE{$(S \neq \emptyset)$}\medskip
\STATE{$\displaystyle d := \arg\min_c \left\{ \left. \frac{R_c^+ - R_c(\mathbf{N})}{R_c^+}\ \right|\ c=1, \ldots, C \right\}$}\label{line:defd}\medskip
\STATE{$\displaystyle j := \arg\min_k \left\{ \left. R_c(\mathbf{N} - \mathbf{1}_k) - R_c^+\ \right|\ k \in S \right\}$}\label{line:defj}\medskip
\IF{$\left( R_c(\mathbf{N} - \mathbf{1}_j) > R_c^+\ \mbox{for any}\ c\right)$}
\STATE{$S := S \setminus \{ j \}$}\hfill\COMMENT{No instance of $\mathcal{W}_j$ can be removed}
\ELSE
\STATE{$\mathbf{N} := \mathbf{N} - \mathbf{1}_j$}
\IF{$\left(N_j = \mathit{Nmin}_j\right)$}
\STATE{$S := S \setminus \{ j \}$}
\ENDIF
\ENDIF
\ENDWHILE
\STATE{\textbf{Return} $\mathbf{N}$}
\end{algorithmic}
\end{small}
\end{algorithm}

\section{Numerical Results}\label{sec:results}

We performed a set of numerical simulation experiments to assess the
effectiveness of~\pippo; results will be described in this section. We
implemented Algorithms~\ref{alg:pippo}, \ref{alg:acquire}
and~\ref{alg:release} using GNU Octave~\cite{eaton:2002}, an
interpreted language for numerical computations.

In the first experiment we considered $K=10$ Web Services and $C=5$
workflow types. Service demands $D_{c k}$ have been randomly
generated, in such a way that class $c$ workflows have service demands
which are uniformly distributed in $[0, c/C]$. Thus, class 1 workflows
have lowest average service demands, while type $C$ workflows have
highest demands. The system has been simulated for $T=200$ discrete
steps $t=1, \ldots, T$; each step corresponds to a time interval of
length $W$, long enough to amortize the reconfiguration costs.

Arrival rates $\boldsymbol{\lambda}(t)$ at step $t$ have been
generated according to a fractal model, starting from a randomly
perturbed sinusoidal pattern to mimic periodic fluctuations. Each
workflow type has a different period.

\begin{figure}[t]
\centering\includegraphics[width=\columnwidth]{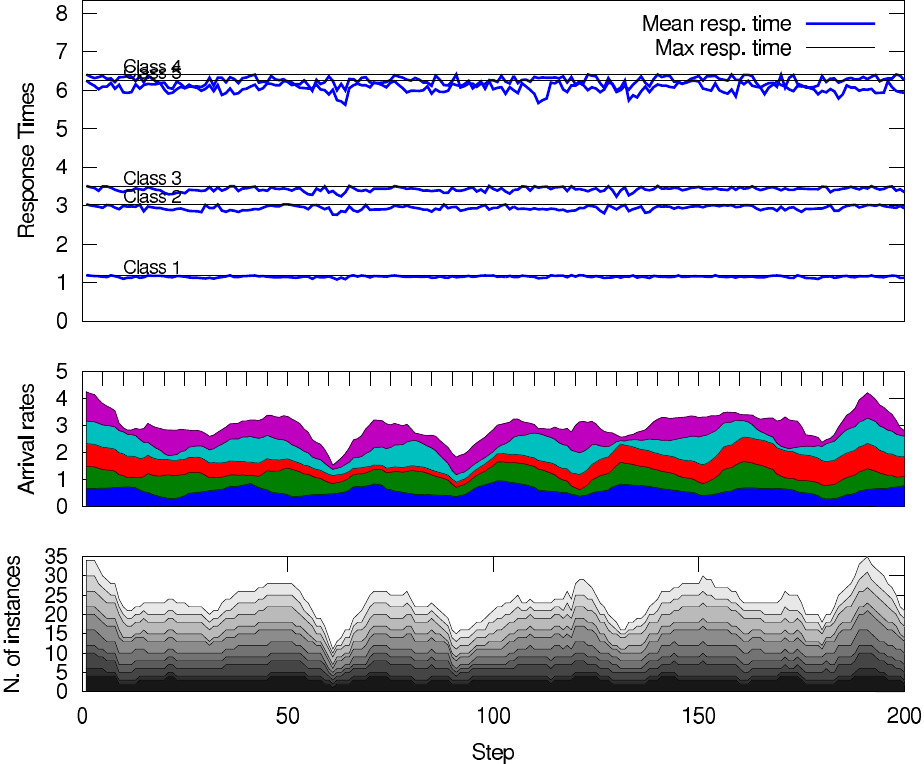}
\caption{Simulation results}\label{fig:example1}
\end{figure}

Figure~\ref{fig:example1} shows the results of the simulation. The top
part of the figure shows the estimated response time $R_c(\mathbf{N})$
(thick lines) and upper limit $R_c^+$ (thin horizontal lines) for each
class $c=1, \ldots, C$. The middle part of the figure shows the
arrival rates $\lambda_c(t)$ for each class $c=1, \ldots, C$; note
that arrival rates have been stacked for clarity, such that the height
of each individual band corresponds to the value $\lambda_c(t)$ from
$c=1$ (bottom) to $c=5$ (top). The total height of the middle graph is
the total arrival rate of all workflow types. Finally, each band of
the bottom part of Figure~\ref{fig:example1} shows the number $N_k$ of
instances of~\ac{WS} $\mathcal{W}_k$, from $k=1$ (bottom) to $k=10$
(top); again, the height of all areas represents the total number of
resources which are allocated at each simulation step. As can be seen,
the number of allocated resources closely follows the workload
pattern.

We performed additional experiments in order to assess the efficiency
of allocations produced by \pippo. In particular, we are interested in
estimating the reduction in the number of allocated instances produced
by \pippo. To do so, we considered different scenarios for all
combinations of $C \in \{10, 15, 20\}$ workflow types and $K \in \{20,
40, 60\}$ Web Services. Each simulation has been executed for $T=200$
steps; everything else (requests arrival rates, service demands) have
been generated as described above.

\begin{table*}[t]
\renewcommand{\arraystretch}{1.3}
\caption{Simulation results for different scenarios}\label{tab:results}
\centering\begin{tabular}{ccccccccccccc}
\hline
&&\multicolumn{2}{c}{Iter. \textsc{Acquire}()} && \multicolumn{2}{c}{Iter. \textsc{Release}()} && \multicolumn{3}{c}{WS Instances (dynamic)}\\
\cline{3-4}\cline{6-7}\cline{9-11}
$C$ & $K$ & \emph{max} & \emph{avg} && \emph{max} & \emph{avg} && \emph{min} & \emph{max} & \emph{tot} & WS Instances (static) & Dynamic/Static \\
\hline
10 & 20 & 14 & 1.30 && 15 & 2.53 && 36 & 127 & 16589 & 25400 & 0.65\\
10 & 40 & 22 & 2.43 && 19 & 3.81 && 76 & 257 & 33103 & 51400 & 0.64\\
10 & 60 & 35 & 3.54 && 35 & 5.12 && 122 & 378 & 50211 & 75600 & 0.66\\
15 & 20 & 10 & 1.27 && 13 & 2.56 && 78 & 178 & 23536 & 35600 & 0.66\\
15 & 40 & 23 & 2.20 && 26 & 3.68 && 138 & 340 & 44843 & 68000 & 0.66\\
15 & 60 & 34 & 3.20 && 44 & 5.04 && 239 & 526 & 68253 & 105200 & 0.65\\
20 & 20 & 9 & 1.19 && 13 & 2.50 && 114 & 206 & 28792 & 41200 & 0.70\\
20 & 40 & 24 & 2.33 && 29 & 4.00 && 215 & 408 & 57723 & 81600 & 0.71\\
20 & 60 & 21 & 3.00 && 30 & 4.89 && 347 & 602 & 86684 & 120400 & 0.72\\
\hline
\end{tabular}
\end{table*}

Results are reported in Table~\ref{tab:results}. Columns labeled $C$
and $K$ show the number of workflow types and Web Services,
respectively. Columns labeled \emph{Iter. \textsc{Acquire}()} contain
the maximum and average number of iterations performed by procedure
\textsc{Acquire}() (Algorithm~\ref{alg:acquire}); columns labeled
\emph{Iter. \textsc{Release}()} contain the same information for
procedure \textsc{Release}() (Algorithm~\ref{alg:release}). Then, we
report the minimum, maximum and total number of resources allocated by
\pippo during the simulation run. Formally, let $S_t$ denote the total
number of~\ac{WS} instances allocated at simulation step $t$; then

\begin{eqnarray*}
\textrm{Min. instances} &=& \min_t \{S_t\}\\
\textrm{Max. instances} &=& \max_t \{S_t\}\\
\textrm{Total instances} &=& \sum_t S_t\\
\end{eqnarray*}

Column labeled \emph{WS Instances (static)} shows the number of
instances that would have been allocated by provisioning for the worst
case scenario; this value is simply $T \times \max_t\{ S_t \}$. The
last column shows the ratio between the total number of~\ac{WS}
instances allocated by \pippo, and the number of instances that would
have been allocated by a static algorithm to satisfy the worst-case
scenario; lower values are better.

The results show that \pippo allocates between 64\%--72\% of the
instances required by the worst-case scenario. As previously observed,
if the~\ac{IaaS} provider charges a fixed price for each instance
allocated at each simulation step, then \pippo allows a consistent
reduction of the total cost, while still maintaining the
negotiated~\ac{SLA}.

\section{Conclusions and Future Works}\label{sec:conclusions}

In this paper we presented \pippo, a QoS-aware algorithm for executing
workflows involving Web Services hosted in a Cloud environment.  The
idea underlying \pippo is to selectively allocate and deallocate Cloud
resources to guarantee that the response time of each class of
workflows is less than a negotiated threshold. The capability of
driving the dynamic resource allocation is achieved though the use of
a feedback control loop. A passive monitor collects information that
is used to identify the minimum number of instances of each~\ac{WS}
which should be allocated to satisfy the response time
constraints. The system performance at different configurations is
estimated using a~\ac{QN} model; the estimates are used to feed a
greedy optimization strategy which produces the new configuration
which is finally applied to the system. Simulation experiments show
that \pippo can effectively react to workload fluctuations by
acquiring/releasing resources as needed.

The methodology proposed in this paper can be improved along several
directions. In particular, in this paper we assumed that all requests
of all classes are evenly distributed across the~\ac{WS} instances.
While this assumption makes the system easier to analyze and
implement, more effective allocations could be produced if we allow
individual workflow classes to be routed to specific~\ac{WS}
instances. This extension would add another level of complexity
to \pippo, which at the moment is under investigation.  We are also
exploring the use of forecasting techniques as a mean to trigger
resource allocation and deallocation proactively.  Finally, we are
working on the implementation of our methodology on a real testbed, to
assess its effectiveness through a more comprehensive set of real
experiments.

\appendix

Let $\curc$ be the current system configuration; let us assume that,
under configuration $\curc$, the observed arrival rates are
$\boldsymbol{\lambda} = (\lambda_1, \ldots, \lambda_C)$ and service
demands are $D_{c k}(\curc)$. Then, for an arbitrary configuration
$\mathbf{N}$, we can combine Equations~\eqref{eq:resptimeck}
and~\eqref{eq:resptime} to get:

\begin{equation}
R_c(\mathbf{N}) = \sum_{k=1}^K N_k \frac{D_{c k}(\mathbf{N})}{1 - U_k(\mathbf{N})}\label{eq:resptime1}
\end{equation}

The current \emph{total} class $c$ service demand on all instances of
$\mathcal{W}_k$ is $M_k D_{c k}(\curc)$, hence we can express service
demands and utilizations of individual instances for an arbitrary
configuration $\mathbf{N}$ as:

\begin{eqnarray}
D_{c k}(\mathbf{N}) &=& \frac{\curv_k}{N_k} D_{c k}(\curc)\label{eq:demand1}\\
U_k(\mathbf{N}) &=& \frac{\curv_k}{N_k} U_k(\curc)\label{eq:utilization1}
\end{eqnarray}

Thus, we can rewrite~\eqref{eq:resptime1} as

\begin{equation}
R_c( \mathbf{N} ) = \sum_{k=1}^K \frac{D_{c k}(\curc) \curv_k N_k}{N_k - U_k(\curc) \curv_k}\label{eq:respestimate}
\end{equation}

\noindent which allows us to estimate the response time
$R_c(\mathbf{N})$ of class $c$ workflows, given information collected
by the monitor for the current configuration $\curc$.

From~\eqref{eq:utilization} and~\eqref{eq:demand1} we get:

\begin{equation}
U_k(\mathbf{N}) = \frac{\curv_k}{N_k} \sum_{c=1}^C \lambda_c D_{c k}(\curc)\label{eq:util2}
\end{equation}

Since by definition the utilization of any~\ac{WS} instance must be
less than one, we can use~\eqref{eq:util2} to define a lower bound on
the number $N_k$ of instances of $\mathcal{W}_k$ as:

\begin{equation}
N_k \geq \curv_k \sum_{c=1}^C \lambda_c D_{c k}(\curc)\label{eq:capconstraint}
\end{equation}

The following lemma can be easily proved:\medskip

\begin{lemma}\label{lemma:monotonically}
The response time function $R_c(\mathbf{N})$ is monotonically
decreasing: for any two configurations $\mathbf{N}'$ and
$\mathbf{N}''$ such that $N'_k \leq N''_k$ for all $k=1, \ldots, K$,
we have that $R_c(\mathbf{N}') \geq R_c(\mathbf{N}'')$
\end{lemma}

\begin{IEEEproof} 
If we extend $R_c(\mathbf{N})$ to be a continuous function, its
partial derivative is
\begin{equation}
\frac{\partial R_c}{\partial N_k} = \frac{- \curv_k^2 U_k(\curc) D_{ck}(\curc)}{\left( N_k - U_k(\curc) \curv_k \right)^2}\label{eq:deriv}
\end{equation}
which is less than zero for any $k$ for which the utilization
$U_k(\curc)$ and service demand $D_{ck}(\curc)$ are nonzero. Hence,
function $R_c(\mathbf{N})$ is decreasing.
\end{IEEEproof}\medskip

Note that, according to Eq.~\eqref{eq:respestimate}, response time
increments are additive. This means that 
$R_c(\mathbf{N}) - R_c(\mathbf{N} + \mathbf{1}_j) =
\boldsymbol{\Delta}_j$ and $R_c(\mathbf{N}) - R_c(\mathbf{N} +
\mathbf{1}_i) = \boldsymbol{\Delta}_i$ imply $ R_c(\mathbf{N})
- R_c(\mathbf{N} + \mathbf{1}_i + \mathbf{1}_j) =
\boldsymbol{\Delta}_i + \boldsymbol{\Delta}_j$

\bibliographystyle{IEEEtran}
\bibliography{cloud_workflows}

\end{document}